# Extracting the Redox Orbitals in Li Battery Materials with High-Resolution X-Ray Compton Scattering Spectroscopy

K. Suzuki[1*], B. Barbiellini[2], Y. Orikasa[3], N. Go[1], H. Sakurai[1], S. Kaprzyk[2,4], M. Itou[5], K. Yamamoto[3], Y. Uchimoto[3], Yung Jui Wang[2,6], H. Hafiz[2], A. Bansil[2] and Y. Sakurai[5]

[1]*Faculty of Science and Technology, Gunma University, Tenjin-cho, Kiryu, Gunma 376-8515, Japan*

[2]*Physics Department, Northeastern University, Boston, Massachusetts 02115, USA*

[3]*Graduate School of Human and Environmental Studies, Kyoto University, Yoshida-nihonmatsu-cho, Sakyo-ku, Kyoto 606-8501, Japan*

[4]*Faculty of Physics and Applied Computer Science, AGH University of Science and Technology, aleja Mickiewicza 30, 30-059 Krakow, Poland*

[5]*Japan Synchrotron Radiation Research Institute, SPring-8, 1-1-1 Kouto, Sayo, Sayo, Hyogo 679-5198, Japan*

[6]*Advanced Light Source, Lawrence, Berkeley National Laboratory, Berkeley, California 94720, USA*

* kosuzuki@gunma-u.ac.jp

**Abstract**

We present an incisive spectroscopic technique for directly probing redox orbitals based on bulk electron momentum density measurements via high-resolution x-ray Compton scattering. Application of our method to spinel $Li_xMn_2O_4$, a lithium ion battery cathode material, is discussed. The orbital involved in the lithium insertion and extraction process is shown to mainly be the oxygen $2p$ orbital. Moreover, the manganese $3d$ states are shown to experience spatial delocalization involving $0.16 \pm 0.05$ electrons per Mn site during the battery operation. Our analysis provides a clear understanding of the fundamental redox process involved in the working of a lithium ion battery.

The doping of oxide electronics is in some ways similar to conventional doping of semiconductors but yields much greater functionality through electronic control of redox phenomena [1] with possible applications to superconductors, spintronics devices, batteries, and solar and fuel cells. This redox paradigm is invoked, for example, when holes are introduced in an oxide. In this case, one can consider the concept of metal-oxygen hybridization for understanding the role of oxygen participation in the redox orbital. This hybridization can be interpreted as an indicator of the electropositive character of the metal with respect to oxygen. A prominent example involves the high $T_c$ superconductors where $Cu^{3+}$ can be visualized as $Cu^{2+}$ with a hole on the oxygen ligand sites [2]. Curiously, here the copper ions become more electronegative than oxygen. Similar effects may occur in lithium ion batteries [3-6]. In fact, significance of holes with strong O $2p$ character was indicated by soft x-ray absorption spectroscopy of $Li_xCoO_2$. Mizokawa *et al.* concluded that the holes play an essential role in the electron conductivity of the $Co^{3+}/Co^{4+}$ mixed valence $CoO_2$ layer [7]. However, the complexity added by electron correlation effects in transition metals modifies the redox mechanism in other battery cathode materials [8].

One of the most interesting cathode materials is the spinel $Li_xMn_2O_4$ because of its low cost and a good record of safety due to the thermodynamic stability of $LiMn_2O_4$ [9]. Like magnetite [10], stoichiometric $LiMn_2O_4$ undergoes a Verwey transition from the cubic phase to a distorted structure below 290 K, which is attributed to the localization of some manganese $3d$ electrons and the associated lattice distortions [11,12]. Unfortunately, this phase transition induces volume changes, leading to the fracture of the electrode and degradation of battery performance with repeated lithium insertion and extraction cycles. In fact, the manganese ions in $LiMn_2O_4$ are reported to have a mixed valence state between $Mn^{3+}$ and $Mn^{4+}$ [13-15] with high spin configurations favoring distortions associated with the Verwey transition. Concerning the orbitals involved in the redox process in $Li_xMn_2O_4$, current interpretations of electronic structure calculations lead to contradictory pictures. Some studies conclude charge transfer from lithium to oxygen ions upon lithium insertion [16,17], while others claim that charge transfer takes place from lithium to manganese $3d$ states [18,19].

In this Letter, we clearly demonstrate the existence of the $O^{2-}/O^{-}$ redox process in $Li_xMn_2O_4$ with x-ray Compton scattering by showing that variation of the hole concentration mostly affects the $2p$ character of oxygen ligand sites. We support this scenario via robust first-principles calculations in which we include subtle disorder effects [20] due to lithium doping and $3d$ electron magnetic moments.

Extracting the character of redox orbitals from inelastic x-ray scattering data is a considerable challenge because it involves delicate differences in spectra taken at different lithium concentrations. Therefore, highly accurate data are required to enable zooming in on weak changes in electron occupancy resulting from lithium insertion or extraction, and to thus identify footprints of redox orbitals therein. Moreover, one needs to measure an observable connected to the one-body reduced density matrix operator, such as the Compton profile $J(p_z)$, obtained from x-ray Compton scattering experiments. Within the impulse approximation [21,22] $J(p_z)$ is given by

$$J(p_z) = \iint \rho(\mathbf{p}) dp_x dp_y, \qquad (1)$$

where $\mathbf{p} = (p_x, p_y, p_z)$ is the electron momentum and $\rho(\mathbf{p})$ is the electron momentum density. The natural orbitals $\Psi_j(\mathbf{r})$ and their occupation numbers $n_j$ can be used to express $\rho(\mathbf{p})$ in term of one-particle states [23,24],

$$\rho(\mathbf{p}) = \sum_j n_j \left| \int \Psi_j(\mathbf{r}) \exp(-i\mathbf{p} \cdot \mathbf{r}) d\mathbf{r} \right|^2 \quad . \tag{2}$$

In this way, the Compton profile directly probes one-particle wave functions. A major advantage of x-ray Compton scattering over other spectroscopies is that it allows a bulk-sensitive measurement of a disordered system under sample conditions of temperature, magnetic field and pressure. It can also probe a sample inside a metal container, which can be penetrated by high-energy x-rays, while other spectroscopies involving incoming or outgoing charged particles (e.g., photoemission) cannot be used. Moreover, the observed electron momentum density is a ground-state property, which allows us to interpret the experimental spectra relatively straightforwardly through first-principles electronic structure calculations without the complications associated with matrix element effects in photoemission [25], positron-annihilation [26], scanning tunneling [27] and other highly resolved spectroscopies. In fact, x-ray Compton scattering has been used as a robust technique for investigating wave-function characters [2,28,29] and Fermi surfaces [30-32] in wide classes of materials. Compton scattering has also been applied to study charge transfers in metal alloys [33-36] and also in Li battery cathode materials [37]. Moreover, the determination of the Mulliken analysis has been very successful to explain the behavior of the Compton profile in Si clathrates [38]. Therefore, these investigations demonstrate how exquisitely sensitive Compton scattering is to charge transfer. Here, we show how this spectroscopy reveals the anion character of the redox orbital which may play a crucial role for the capacity of the Li batteries.

The present measurements were performed at room temperature using the x-ray spectrometer for Compton scattering at the BL08W beamline of SPring-8. The incident x-ray energy was 115 keV and the scattering angle was 165 degrees. The energy spectrum of Compton scattered x-rays was converted to the Compton profile [21], which was then normalized to the total number of valence electrons after the core electron contribution was subtracted. Each profile was corrected to address absorption contributions, analyzer and detector efficiencies, multiple scattering contributions, and x-ray backgrounds. Here, we employ the relativistic cross section for x-ray Compton scattering [39]. The overall momentum resolution was 0.10 atomic units (a.u.). The valence Compton profiles from $Li_xMn_2O_4$ for three different Li concentrations are given in the Supplemental Material [40].

Polycrystalline samples of $Li_xMn_2O_4$ ($x$=0.496, 1.079, and 1.233) were prepared by chemical lithium extraction. The compositions were determined by inductively coupled plasma (ICP) measurements. X-ray powder diffraction analyses confirmed a single spinel phase with $Fd\bar{3}m$ space group with the lattice constants of the three samples following Vegard's law. Pellets of these samples, 10 mm in diameter and 2 mm in thickness, were produced with a cold isostatic press.

In order to extract the redox orbitals, we consider the difference in the Compton profiles ($\Delta J(p_z)$) between two samples with different lithium concentrations as shown in Fig. 1(a). This subtraction enables us to zoom in on changes in electron occupancy near the Fermi level associated with lithium insertion or extraction by eliminating contributions of the core as well as the irrelevant valence electrons far away from the Fermi level [2]. As Eq. (2) shows, the electron momentum density is given by the squared modulus of the occupied momentum-space wave functions (connected to the position-space wave functions through a Fourier transform). Note that, in general, each atomic orbital yields its own distinct radial dependence in the electron momentum density determined by the related spherical Bessel function, which behaves as $p^l$ at small momenta $p$, where $l$ is the orbital quantum number [43]. Therefore, an oxygen $2p$ orbital contributes to the electron momentum density at low momenta, whereas the contribution of manganese $3d$ orbitals extends to high momenta. This implies that the Compton profile is narrow for oxygen $2p$ states while it is broad for manganese $3d$ states as illustrated in Fig. 1(b). This tendency carries over to a molecular state [44] and applies also to the solid state [45].

We now focus on the identification of redox orbitals whose properties are changed by lithium insertion and extraction. Electronic structure of the spinel $Li_xMn_2O_4$ is mainly determined by the $MnO_6$ octahedron with electronic states primarily of oxygen $2p$ and manganese $3d$ character. The strong bonding between the oxygen $2p$ and manganese $3d$ states produces a host structure for inserting or extracting lithium ions reversibly without a substantial modification of the host structure [17,18]. Lithium $2s$ electrons are transferred to or removed back from redox states located near the Fermi level. It is natural therefore to compare the experimental spectra with calculated Compton profiles for atomic oxygen and manganese orbitals. As shown in Fig. 1(b), the manganese $3d$ orbital possesses a much broader profile than oxygen $2p$, and the oxygen $2p$ profile shows a much better agreement with the experiment with respect to momenta below $p_z \sim 1$ a.u. [Fig. 1(a)].

A model of electronic states originating in oxygen $2p$ orbitals captures the nature of redox orbitals in spinel $Li_xMn_2O_4$. This simple atomic picture describes the main features, even if this scheme does not include the itinerant behavior of the solid-state. The Bloch character of the one-particle states is of course properly described in band structure calculations. Accordingly, for a more realistic interpretation of the spectra, we have calculated Compton profiles for $Li_xMn_2O_4$ for $x$=0.0, 0.25, 0.5, 0.75, 1.0, 1.2, and 1.5, based on first-principles Korringa-Kohn-Rostoker coherent-potential-approximation (KKR CPA) computations within the framework of the local spin density approximation [46,47]. In these calculations, we treat the spinel structure of $Li_xMn_2O_4$ in which lithium and empty sites are randomly occupied by lithium atoms and vacancies [48] with appropriate probabilities. Our results for pristine $LiMn_2O_4$ are in excellent agreement with the full potential calculations obtained within the Wien2k code [49], demonstrating the high quality of our basis set. The orientation of spin at each atomic site has also been treated randomly to simulate a spin-glass-like behavior produced by the geometrical frustration of the antiferromagnetic ordering in the pyrochlore network in the spinel structure, combined with the substitution disorder related to lithium doping.

As shown in Fig. 1(a), the theoretical difference Compton profile between $x$=0.5 and $x$=1.0, $x$=1.0 and $x$=1.2 shows an excellent agreement with experiment, indicating that the present calculations capture correctly the key features of the electronic structure and the associated electron momentum density distribution. These electronic structure calculations thus provide a reliable tool for identifying the redox orbitals.

Figure 2(a) shows clearly that the number of electrons in the interstitial region increases as the lithium concentration increases, while the number of electrons at the manganese sites is almost constant. The number of electrons within the oxygen muffin-tin radius is also constant, but some of the extra charge in the interstitial region belongs to the oxygens since the oxygen ionic radius is larger than the muffin-tin radius. This can be confirmed through an analysis of partial densities of states (partial DOS) associated with just the muffin-tin spheres compared to the total contribution from all the electrons (total DOS), including those in the interstitial region [Fig. 2(b)]. These comparisons mostly differ in the region where the oxygen 2$p$ dominates. We also carried out a Mulliken population analysis using the CRYSTAL09 code [50,51], which shows clearly that 0.96 electrons of lithium 2$s$ go to oxygen 2$p$ orbitals when the lithium concentration goes from 0 to 1.

Interestingly, our analysis shows that, for Li$_x$Mn$_2$O$_4$ with 0< $x$ <1, the redox orbitals display a large delocalized oxygen 2$p$ character, indicating high anionic redox activity. The delocalization of redox orbitals makes electron transport possible in the cathode. Therefore, this effect is an important condition for facile redox reactions inside the cathode materials. The importance of oxygen character has also been noted by Mizokawa *et al.* in a soft x-ray absorption study of Li$_x$CoO$_2$ [7], and opens a possible pathway for developing higher capacity cathode materials [52,53]. The present approach combining x-ray Compton scattering measurements with first-principles modeling provides a unique method for quantitatively extracting and monitoring anionic character in redox orbitals.

Although oxygen 2$p$ states play a dominant role in the redox process, the manganese 3$d$ states control the magnetic properties. KKR CPA calculations in Fig. 2(c) predict that the effective magnetic moment of manganese atoms increases as the lithium concentration increases if the system is described by a spin-glass-like behavior with randomly oriented manganese effective moments. Strengthening of the manganese effective magnetic moment with lithium insertion is consistent with muon spin rotation experiments [54] but contradicts other electronic structure calculations performed for ferromagnetic or antiferromagnetic spin configurations [19].

Interestingly, the observed negative part between 1 a.u. and 2 a.u. in the difference Compton profile [see Fig. 1(a)] can be accounted for if some $d$ electrons are transferred from localized to less localized $d$ states as shown in Fig. 3(a). We can estimate that the number of electrons in the negative part of the difference Compton profile is 0.16 ± 0.05 per lithium atom, indicating that if one electron goes from the valence of lithium to a delocalized O $2p$ orbital, this transfer induces a wave-function modification of about 0.16 electrons in the manganese $3d$ shell. The small uncertainty of about 0.05 electrons demonstrates that the negative excursion is a robust feature in momentum space. A similar error analysis has been used to analyze wave-function localization properties in manganites from Compton profiles [29]. Nevertheless, the manganese $3d$ partial DOS in Fig. 3(b) shows that the total number of $d$ electrons does not change much at the manganese site, suggesting that a strict definition of ionic $d$ states is not very meaningful in the present case. Moreover, when $x > 1$, the negative part of the Compton profile difference ($x=1.233$) – ($x=1.079$) disappears since the number of electrons in this region is estimated to be zero within the present error bars. In fact, according to the KKR CPA calculations corresponding to the Compton profile difference ($x=1.2$) – ($x=1.0$), the negative excursion is absent as shown in the inset of Fig. 1(a). Therefore, in this regime, the behavior of the Mn $3d$ wavefunctions no longer follows the delocalization pattern depicted in Fig. 3(c).

In conclusion, our study demonstrates the suitability of high-resolution x-ray Compton scattering as a direct and bulk-sensitive probe of the active orbitals in the $Li_xMn_2O_4$ redox process. The x-ray Compton spectra provide a robust signature (insensitive to defects, surfaces, and impurities) and an image (in momentum space) of the electronic ground state. Although we have discussed the lithiation of $Li_xMn_2O_4$ in the present study, our approach is applicable to other battery materials such as $LiCoO_2$ [7] and LiFePO$_4$ [8]. Thus, our quantum mechanical approach paves the way for an advanced characterization of lithium ion batteries, in which the redox orbital becomes the focus of the materials design and engineering.

Acknowledgements

K.S. was supported by a Grant-in-Aid for Young Scientists (B) (No. 24750065) from the Ministry of Education, Culture, Sports, Science, and Technology (MEXT), Japan, and the work at JASRI was partially supported by the Japan Science and Technology Agency. Compton scattering experiments were performed with the approval of JASRI (Proposal No. 2011A1869, No. 2011B2004, and No. 2012B1470). The work at Northeastern University was supported by the U.S. Department of Energy, Office of Science, Basic Energy Sciences Grant No. DE-FG02-07ER46352, and benefited from Northeastern University's Advanced Scientific Computation Center (ASCC), and the allocation of time at the NERSC supercomputing center through DOE Grant No. DE-AC02-05CH11231. S.K. was supported by the Polish National Science Center (NCN) under Grant No. DEC-2011/02/A/ST3/00124.

References

[1]  G. Pacchioni and H. Freund, Chem. Rev. **113**, 4035 (2013).

[2]  Y. Sakurai, *et al*., Science **332**, 698 (2011).

[3]  M. Armand and J.-M. Tarascon, Nature (London) **451**, 652 (2008).

[4]  J. B. Goodenough and Y. Kim, Chem. Mater. **22**, 587 (2010).

[5]  J. Tollefson, Nature (London) **456**, 436 (2008).

[6]  J.-M. Tarascon and M. Armand, Nature (London) **414**, 359 (2001).

[7]  T. Mizokawa, Y. Wakisaka, T. Sudayama, C. Iwai, K. Miyoshi, J. Takeuchi, H. Wadati, D. G. Hawthorn, T. Z. Regier, and G. A. Sawatzky, Phys. Rev. Lett. **111**, 056404 (2013).

[8]  X. Liu, *et al*., J. Am. Chem. Soc. **134**, 13708 (2012).

[9]  M. M. Thackeray, Prog. Solid State. Chem. **25**, 1 (1997).

[10] J. Garcia and G. Subias, J. Phys. Condens. Matter **16**, R145 (2004).

[11] A. Yamada and M. Tanaka, Mater. Res. Bull. **30**, 715 (1995).

[12]   H. Yamaguchi, A. Yamada, and H. Uwe, Phys. Rev. B **58**, 8 (1998).

[13]   C. R. Horne, U. Bergmann, M. M. Grush, R. C. C. Perera, D. L. Ederer, T. A. Callcott, E. J. Cairns, and S. P. Cramer, J. Phys. Chem. B **104**, 9587 (2000).

[14]   J. Rodriguez-Carvajal, G. Rousse, C. Masquelier, and M. Hervieu, Phys. Rev. Lett. **81**, 4660 (1998).

[15]   Y. Ein-Eli, R. C. Vrian, W. Wen, and S. Mukerjee, Electrochem. Acta **50**, 1931 (2005).

[16]   M. K. Aydinol and G. Ceder, J. Electrochem. Soc. **144**, 3832 (1997).

[17]   Y. Liu, T. Fujiwara, H. Yukawa, and M. Morinaga, Solid State Ionics **126**, 209 (1999).

[18]   H. Berg, K. Göransson, B. Noläng, and J. O. Thomas, J. Mater. Chem. **9**, 2813 (1999).

[19]   G. E. Grechnev, R. Ahuja, B. Johansson, and O. Eriksson, Phys. Rev. B **65**, 174408 (2002).

[20] A. Bansil, R. S. Rao, P. E. Mijnarends, and L. Schwartz, Phys. Rev. B **23**, 3608 (1981); P. E. Mijnarends and A. Bansil, Phys. Rev. B **13**, 2381 (1976).

[21] W. Schülke, in *X-Ray Compton Scattering*, edited by M. J. Cooper, P. E. Mijnarends, N. Shiotani, N. Sakai, and A. Bansil (Oxford University Press, Oxford, 2004), p. 31-39.

[22] I. G. Kaplan, B. Barbiellini, and A. Bansil, Phys. Rev. B **68**, 235104 (2003).

[23] B. Barbiellini, J. Phys. Chem. Solids **61**, 341 (2000).

[24] B. Barbiellini and A. Bansil, J. Phys. Chem. Solids **62**, 2181 (2001).

[25] S. Sahrakorpi, M. Lindroos, R. S. Markiewicz, and A. Bansil, Phys. Rev. Lett. **95**, 157601 (2005); A. Bansil, M. Lindroos, S. Sahrakorpi, and R. S. Markiewicz, Phys. Rev. B **71**, 012503 (2005).

[26] L. C. Smedskjaer, A. Bansil, U. Welp, Y. Fang, and K. G. Bailey, J. Phys. Chem. Solids **52**, 1541 (1991); J. C. Campuzano, L. C. Smedskjaer, R. Benedek, G. Jennings, and A. Bansil, Phys. Rev. B **43**, 2788 (1991).

[27] J. Nieminen, H. Lin, R. S. Markiewicz, and A. Bansil, Phys. Rev. Lett. **102**, 037001 (2009).

[28]     A. Koizumi, S. Miyaki, Y. Kakutani, H. Koizumi, N. Hiraoka, K. Makoshi, N. Sakai, K. Hirota, and Y. Murakami, Phys. Rev. Lett. **86**, 5589 (2001).

[29]     B. Barbiellini, A. Koizumi, P. E. Mijnarends, W. Al-Sawai, Hsin Lin, T. Nagao, K. Hirota, M. Itou, Y. Sakurai, and A. Bansil, Phys. Rev. Lett. **102**, 206402 (2009).

[30]     S. B. Dugdale, R. J. Watts, J. Laverock, Zs. Major, M. A. Alam, M. Samsel-Czekala, G. Kontrym-Sznajd, Y. Sakurai, M. Itou, and D. Fort, Phys. Rev. Lett. **96**, 046406 (2006).

[31]     N. Hiraoka, T. Buslaps, V. Honkimäki, J. Ahmad, and H. Uwe, Phys. Rev. B **75**, 121101(R) (2007).

[32]     C. Utfeld, *et al*., Phys. Rev. B **81**, 064509 (2010).

[33]     S. Manninen, B. K. Sharma, T. Paakkari, S. Rundqvist, and M. W. Richardson, Phys. Status. Solidi B **107**, 749 (1981).

[34]     E. Zukowski, L. Dobrzynski, M. J. Cooper, D. N. Timms, R. S. Holt, and J. Latuszkiewicz, J. Phys.: Cond. Mat. **2**, 6315 (1990).

[35]     A. Andrejczuki, L. Dobrzynski, E. Zukowski, M. J. Cooper, S. Hamouda, and J. Latuszkiewicz, J. Phys.: Cond. Mat. **4**, 2735 (1992).

[36]   J. Kwiatkowska, B. Barbiellini, S. Kaprzyk, A. Bansil, H. Kawata, and N. Shiotani, Phys. Rev. Lett. **96**, 186403 (2006).

[37]   S. Chabaud, Ch. Bellin, F. Mauri, G. Loupias, S. Rabii, L. Croguennec, C. Pouillerie, C. Delmas, and Th. Buslaps, J. Phys. Chem. Solids **65**, 241 (2004).

[38]   M. Volmer, C. Sternemann, J. S. Tse, T. Buslaps, N. Hiraoka, C. L. Bull, J. Gryko, P. F. McMillan, M. Paulus, and M. Tolan, Phys. Rev. B **76**, 233104 (2007).

[39]   P. Holm, Phys. Rev. A **37**, 3706 (1988).

[40]   See Supplemental Material at [URL], which includes Refs. [41,42], for Compton profiles of $Li_xMn_2O_4$ ($x$ = 0.496, 1.079 and 1.233).

[41]   N. Sakai, J. Phys. Soc. Jpn. **56**, 2477 (1987).

[42]   F. Biggs, L. B. Mendelsohn, and J. B. Mann, At. Data Nucl. Data Tables **16**, 201 (1975).

[43]   P. E. Mijnarends, Physica (Amsterdam) **63A**, 235 (1973).

[44]   W. Weyrich, P. Pattison, and B. G. Williams, Chem. Phys. **41**, 271 (1979).

[45]     R. Harthoorn and P. E. Mijnarends, J. Phys. F **8**, 1147 (1978).

[46]     A. Bansil, B. Barbiellini, S. Kaprzyk, and P. E. Mijnarends, J. Phys. Chem. Solids **62**, 2191 (2001).

[47]     A. Bansil, S. Kaprzyk, P. E. Mijnarends, and J. Tobola, Phys. Rev. B **60**, 13396 (1999).

[48]     S. N. Khanna, A. K. Ibrahim, S. W. McKnight, and A. Bansil, Solid State Commun. **55**, 223 (1985).

[49]     P. Blaha, K. Schwarz, G. K. H. Madsen, D. Kvasnicka, and J. Luitz, Technische Universität Wien, 2001; http://www.wien2k.at.

[50]     R. Dovesi, R. Orlando, B. Civalleri, C. Roetti, V. R. Saunders, and C. M. Zicovich-Wilson, Z. Kristallogr. **220**, 571 (2005)

[51]     R. Dovesi *et al*., *CRYSTAL09 User's Manual* (University of Torino, Torino, 2009).

[52]     M. Sathiya, G. Rousse, K. Ramesha, C. P. Laisa, H. Vezin, M. T. Sougrati, M-L. Doublet, D. Foix, D. Gonbeau, W. Walker *et al*., Nat. Mater. **12**, 827 (2013).

[53]     M. Sathiya, K. Ramesha, G. Rousse, D. Foix, D. Gonbeau, A. S. Prakash, M. L. Doublet, K. Hemalatha, and J.-M. Tarascon, Chem. Mater. **25**, 1121 (2013).

[54]     K. Mukai, J. Sugiyama, K. Kamazawa, Y. Ikedo, D. Andreica, and A. Amato, J. Solid State Chem. **184**, 1096 (2011).

Figure captions

FIG. 1 Electron momentum density distributions of the redox orbitals in Li$_x$Mn$_2$O$_4$. (a) Difference of Compton profiles for two different values of $x$, given by solid black dots for $x$=1.079 and $x$=0.496. $\Delta J(p_z)$, allows zooming in on the electron momentum density distribution of the redox orbitals. Error bars give standard deviation with respect to x-ray photon counts at momenta $p_z$. Experimental distribution is reproduced by the theoretical difference Compton profile (solid red line). Inset: difference profile between $x$=1.233 and $x$=1.079, together with the corresponding theoretical results. Individual experimental profiles for different $x$ values are given in Supplemental Materials [40]. (b) Electron momentum density distributions (Compton profiles) of atomic O 2$p$ and Mn 3$d$ orbitals, showing that O 2$p$ orbital is substantially narrower than Mn 3$d$. a.u. denotes atomic units.

FIG. 2 Theoretically derived characteristics of the redox orbitals. (a) Number of valence electrons on O sites, Mn sites, and the interstitial region as a function of Li concentration $x$. Number of electrons in interstitials ($N_{int}$, green squares, right-hand scale) increases as Li $x$ increases, although the number at Mn sites ($N_{Mn}$, red filled circles, left-hand scale) is almost constant. The number of electrons at O sites ($N_O$, blue triangles, right-hand scale) increases slightly as $x$ increases. (b) Density of states (DOS) for LiMn$_2$O$_4$. Partial DOS contributed by muffin-tin spheres (magenta curve) and total DOS (black curve) differ mostly in the energy region below 0.4 Ry, where O 2$p$ dominates. (c) Effective Mn magnetic moment $m$ as a function of $x$. We see that $m$

increases monotonically as $x$ increases, as a result of electron transfer in the Mn $3d$ shell from down-spin to up-spin electrons.

FIG. 3 Effect of Mn $3d$ localization on electron momentum density distribution. (a) Upper panel: hydrogenlike atomic orbital calculations simulate the effect of Mn $3d$ localization on electron momentum density distribution (Compton profile), where the effective charge ($Z_{\text{eff}}$) is 7 for Li$_x$Mn$_2$O$_4$ (red curve) and 7.5 for Mn$_2$O$_4$ (blue curve). Lower panel: the difference between Li$_x$Mn$_2$O$_4$ and Mn$_2$O$_4$ reproduces the negative part between $p_z$=1.5 a.u. and $p_z$=4 a.u. observed in experiment [Fig. 1(a)]. (b) Mn $3d$ partial DOS for Mn$_2$O$_4$ and LiMn$_2$O$_4$ from KKR CPA calculations. The peaks of the Mn $3d$ partial DOS become more atomic-like as the Li concentration increases. (c) Radial wave function of the Mn $3d$ orbital. Insertion of Li induces a delocalization of the $d$ electrons in real space.

FIG. 1

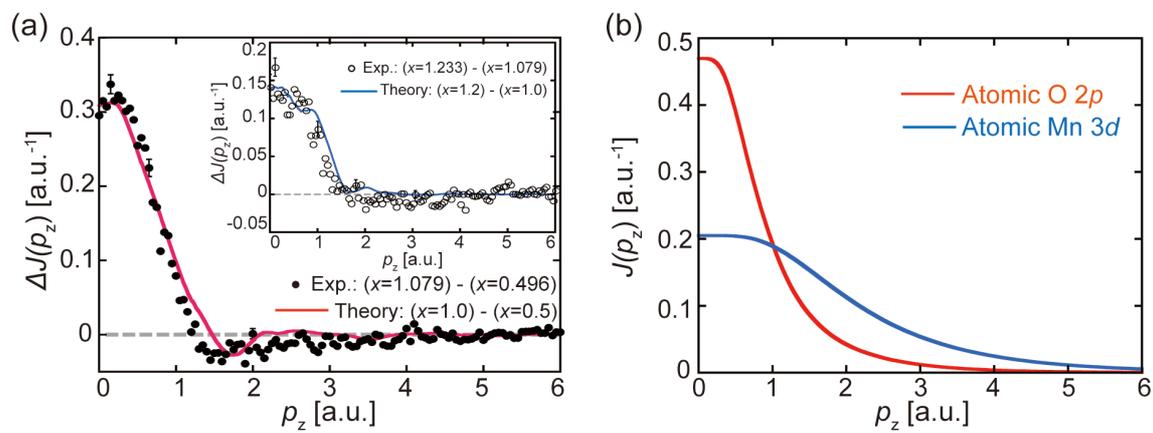

FIG. 2

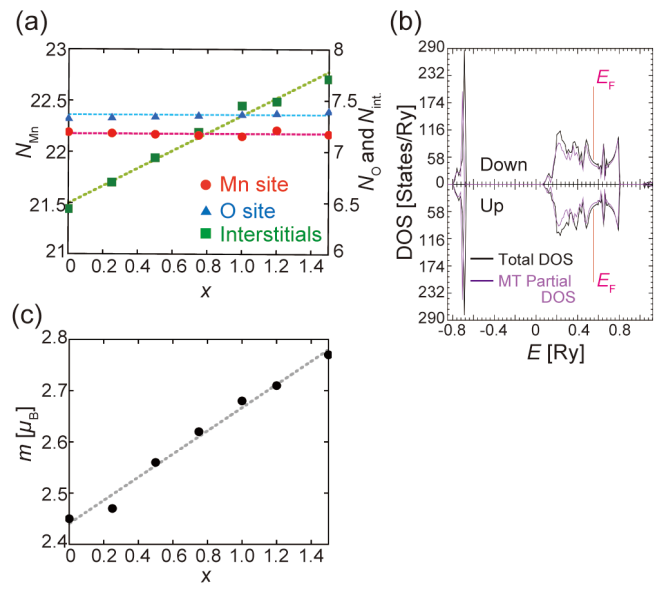

FIG. 3

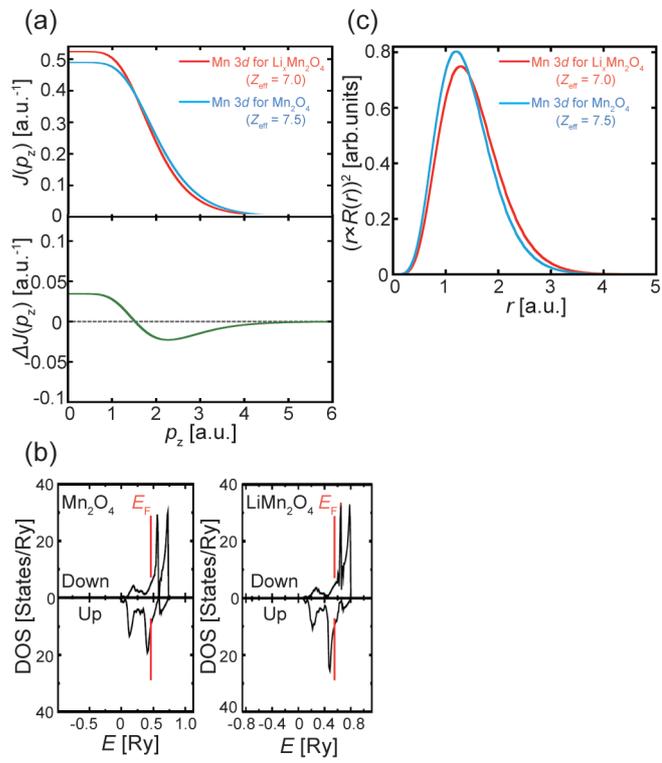